\documentclass[aps,prd,groupedaddress,showpacs]{revtex4}
\usepackage{amsmath}
\usepackage[parfill]{parskip}    % Activate to begin paragraphs with an empty line rather than an indent
\usepackage{float}
\usepackage{caption}
\usepackage{subcaption}
\usepackage{graphicx}
\usepackage{bbm}
\usepackage{graphicx}
\usepackage{amssymb}
\usepackage{epstopdf}
\usepackage{color}
\usepackage{ulem}
\usepackage{graphicx}
\usepackage{pdfpages}
\usepackage[colorlinks=true,citecolor=blue,urlcolor=magenta,breaklinks]{hyperref}
\begin{document}

\title{ $2+1$ Einstein--Klein--Gordon black holes by gravitational decoupling}

\author{Pio J. Arias}
\email{pio.arias@ciens.ucv.ve}
\affiliation{Centro de F\'isica Te\'orica y Computacional,\\ Escuela de F\'isica, Universidad Central de Venezuela. \\}

\author{Pedro \surname{Bargueño}}
\email{pedro.bargueno@ua.es}
\affiliation{Departamento de F\'{i}sica Aplicada, Universidad de Alicante, Campus de San Vicente del Raspeig, E-03690 Alicante, Spain}

\author{Ernesto Contreras}
\email{econtreras@usfq.edu.ec}
\affiliation{Departamento de F\'isica, Colegio de Ciencias e Ingenier\'ia, Universidad San Francisco de Quito,  Quito, Ecuador\\}

\author{Ernesto Fuenmayor}
\email{ernesto.fuenmayor@ciens.ucv.ve}
\affiliation{Centro de F\'isica Te\'orica y Computacional,\\ Escuela de F\'isica, Facultad de Ciencias, Universidad Central de Venezuela, Caracas 1050, Venezuela\\}

%%%%%%%%%%%%%%%%%%%%%%%%%%%%%%%%%%
\begin{abstract}
In this work we study the 2+1- Einstein--Klein--Gordon system in the framework of Gravitational Decoupling. We associate the generic matter decoupling sector with a real scalar field so we can obtain a constraint which allows to close the system of differential equations. The constraint corresponds to a differential equation involving the decoupling functions and the metric of the seed sector and will be independent of the scalar field itself. We show that when the equation admits analytical solutions, the scalar field and the self--interacting potential can be obtained straightforwardly. We found that, in the cases under consideration, it is possible to express the potential as an explicit function of the scalar field only for certain particular cases corresponding to limiting values of the parameters involved.
\\
\\
{\bf keywords}: black holes; gravitational decoupling; minimal geometric deformation
\end{abstract}
%%%%%%%%%%%%%%%%%%%%%%%%%%%%%%%%%%%%%%%%%%

\maketitle

%%%%%%%%%%%%%%%%%%%%%%%%%%%%%%%%%%%%
\section{Introduction}\label{intro}

As a theory of gravitation for spacetime, Einstein’s theory of General Relativity is considered one of the most successful theories in sciences with a wide variety of predictions, many of them in cosmology, interpreting the dynamics of the universe. Despite the incredible success of the theory, recently confirmed by the detection of gravitational waves \cite{Abbott4,Abbott5} and the reconstruction of the shadow of a supermassive black hole \cite{Akiyama1,Akiyama2}, some alternatives approaches are still under considerations given that they could play and important role in some scenarios in which General Relativity could fail. For example, it is a well--known fact that modified gravity theories usually admit more gravitational wave polarizations, and, in turn, different interferometric response functions, with respect to standard General Relativity (see \cite{reff3}, for example). Besides, its geometric character implies that quantizing gravity means quantizing spacetime itself and we do not know what this means. Therefore, it seems natural to look for simpler models that share the important conceptual features of General Relativity and many of the fundamental issues of quantum gravity. In this sense, General Relativity in 2+1 dimensions is of importance \cite{Carlip1998}. Despite 2+1-dimensional gravity has no propagating gravitational degrees of freedom, the seminal discovery of the Ba\~nados-Teitelboim-Zanelli (BTZ) solution \cite{Banados1992,Banados1993} paved the way towards establishing a complex catalog of exact solutions due to non--trivial interactions between gravitational and other fundamental fields. So, three dimensional gravity has resulted a fertile ground in which non--trivial interactions between gravitational and fundamental fields produce exact analytical solutions. The reasons are many, but most of them are simply to get some insights about how to face apparent unsolvable issues in  $3+1$--dimensional models.\\

There is a vast amount of literature about solutions in $2+1$ dimensions in presence of matter and fields, for instance, point particle solutions, perfect fluids, cosmological spacetimes, dilatons, inflatons, and stringy solutions (see \cite{Garcia-Diaz:2017cpv} and references there in). Similarly, these systems have been the preferable arena for toy models in alternative theories as in scale--dependent gravity where the matter fields are associated to non--linear electrodynamics \cite{sd2}. 
However, probably the most interesting problems arise when the source corresponds to certain fundamental field, as for example a scalar field in the framework of Einstein--Klein--Gordon (EKG) system. The EKG equations, representing a scalar field
coupled to gravity, have regular, static and stable solutions, the so-called boson stars. Due to their characteristics of stable and static nature they may have astrophysical relevance \cite{BS3,BS4,BS2,eso}. Either they are related to Dark Matter structures, in the context of Scalar Dark Matter models or, in general, they could be formed by any of the various scalar fields that the extensions to the Standard Model of particles or unification theories predict. \\

Scalar fields are interesting in its own right. The discovery of the Higgs field has opened the possibility for the existence of other scalar fields, such as quintessence which could be responsible for the universe dynamics \cite{q1,q2} or the inflaton \cite{i1,i2}, which plays a key role in the theory of inflation. In any case, scalar fields provide a simple toy model for understanding the behaviour of more general fields in more complicated scenarios. In the cosmological context $2 + 1$ physics and scalar fields has a venerable history. There are scalar field solutions related with Friedmann–Robertson–Walker cosmologies (in flat or curved) spaces, also scalar field solutions for barotropic and linear equations of state coupled with perfect or anisotropic fluids \cite{Garcia-Diaz:2017cpv}. In addition, black hole solutions for minimally and non-minimally coupled scalar fields have been constructed \cite{Martinez1996,Henneaux2002,Correa2012,Karakasis2021}. Very recently, a plethora of new analytic black holes and globally regular three-dimensional horizonless spacetimes have been reported \cite{Bueno2021}. Specifically, there have been several interesting advances in theories considering stationary rotating scalar field solutions in $2+1$ dimensions, using Schwarzschild coordinates frame, for a static cyclic symmetric metric coupled minimally to a dilaton in the presence of an exponential potential. In this case, it is also possible or not, to introduce a Maxwell field \cite{CHM2,CH}. In most of these scenarios in which scalar fields play a primary role, the strategy is always the same: the assumption of the profile of the self--interacting potential seems to be an obligated requirement.\\

In this work we consider the 2+1 EKG system with the aim to
solve for both the scalar field and the self--interacting potential without assuming any ansatz as it is usually done in the literature. To this end, we identify the generic source which arise in the framework of Gravitational Decoupling (GD) \cite{ovalle2017} by the Minimal Geometric Deformation (MGD) approach with the matter sector associated with the scalar field. It is worth mentioning that this has been done in $3+1$ dimensions in \cite{Ovalle:2018ans} (for implementation of MGD in $3+1$ and $2+1$ dimensional spacetimes see \cite{Ovalle:2016pwp,rocha2017a,casadio2017a,estrada2018,fernandez2018,Contreras2019, Ovalle:ext,contrerasextended,lh2019,ovalle2019a,tello2019c,sharif20b,jorgeLibro,Heras:2021xxz} and references therein.).
\\

This work is organized as follows. In the next section we briefly review the GD by MGD in $2+1$ dimensions with cosmological constant. In section \ref{EKG} we introduce the EKG system and identify the associated matter sector with the decoupling source of well known models, namely, the BTZ black hole (BH) and the 2+1 BH with Coulomb field in the framework of the minimal deformation. Section \ref{ec} is devoted to the 
discussion of the extended MGD corresponding case, while final comments and conclusions are given in the last section.

\section{Gravitational decoupling in $2+1$ dimensional space--time with  cosmological term}\label{mgd}

Before entering into the details of the GD by MGD, here we will briefly outline the main features of the method. The GD can be implemented in two different ways. In its reduced form, it implies the so called MGD approach (where only one metric function is deformed) in which the Einstein's equations can be decoupled in two sets: a system of Einstein's equations whose solution is known (which is used as a seed solution) and other {\it decoupling sector} which corresponds to Einstein's field equations in $2+1$ dimensions and quasi-Einstein field equations in $3+1$. The MGD-decoupling works as long as the sources do not exchange energy–momentum among them, which further clarifies that, in this case, their interaction is purely gravitational. In the extended case, the implementation of the GD implies the deformation of both metric potentials, $g_{tt}$ and $g_{rr}$ and, contrary to what happens with MGD, both sources can be successfully decoupled as long as there is exchange of energy between them.
\\
\\
Let us now go into the details of the method.

Let us consider the Einstein field equations
\begin{eqnarray}\label{einsorig}
G_{\mu\nu}+\Lambda g_{\mu\nu}=R_{\mu\nu}-\frac{1}{2}R g_{\mu\nu}+\Lambda g_{\mu\nu}=\kappa^{2}T_{\mu\nu},
\end{eqnarray}
where $G_{\mu\nu}$ corresponds to the Einstein tensor, $R_{\mu\nu}$ and $R$ are the Ricci tensor and the scalar curvature respectively, $T_{\mu\nu}$ is the energy–momentum tensor, that by virtue of the Bianchi identity, fulfills the energy conservation condition $T^{\mu}_{\nu;\mu}=0$, $\Lambda$ is the cosmological constant, $\kappa^{2}$ is a coupling constant, that in three dimensions, is measured in units of inverse of mass and Greek indices are used to label the three--dimensional space--time coordinates, namely, $\mu,\nu,\rho,..=0,1,2$. Now, let us assume that the total energy--momentum tensor can be decomposed as
\begin{eqnarray}\label{total}
T_{\mu\nu}=T_{\mu\nu}^{(s)}+\theta_{\mu\nu},
\end{eqnarray}
where $T^{\mu(s)}_{\nu}=diag(-\rho,p_{r},p_{\perp})$ and $\theta^{\mu}_{\nu}=diag(-\rho^{\theta},p_{r}^{\theta},p_{\perp}^{\theta})$ are sources with a different nature. For example, $T^{(s)}_{\mu\nu}$ could represent the energy momentum tensor of a generic anisotropic fluid and $\theta_{\mu\nu}$ could encode the information of some fundamental field, namely scalar, vector or tensor field. Of course, given the non-linearity of Einstein equations,  Eq. (\ref{total}) does not necessarily imply
that Einstein tensor can be written as
\begin{eqnarray}
G_{\mu\nu}=G_{\mu\nu}^{(s)}+G_{\mu\nu}^{\theta}.
\end{eqnarray}
However, contrary which is broadly believed, such a decomposition can be achieved in a particular case in circularly symmetric space--times as we shall explain in what follows. Let us start with a line element
\begin{eqnarray}\label{le}
ds^{2}=-e^{\nu}dt^{2}+e^{\lambda}dr^{2}+r^{2}d\phi^{2},
\end{eqnarray}
where $\nu$ and $\lambda$ are functions of the radial coordinate, $r$, only. Now, replacing (\ref{total}) and (\ref{le}) in (\ref{einsorig}) we see that the metric must satisfy the Einstein equations, which in terms of the two sources explicitly leads to,
\begin{eqnarray}
\kappa ^2 \tilde{\rho}&=&-\Lambda+\frac{e^{-\lambda} \lambda'}{2 r}\label{eins1}\\
\kappa ^2 \tilde{p}_{r}&=&\Lambda+\frac{e^{-\lambda} \nu '}{2 r}\label{eins2}\\
\kappa ^2 \tilde{p}_{\perp}&=&\Lambda+\frac{1}{4} e^{-\lambda} \left(-\lambda ' \nu '+2 \nu ''+\nu '^2\right)\label{eins3},
\end{eqnarray}
where the prime denotes derivation with respect to the radial coordinate and we have defined
\begin{eqnarray}
\tilde{\rho}&\equiv&\rho+\rho^{\theta},\label{rot}\\
\tilde{p}_{r}&\equiv&p_{r}+p_{r}^{\theta},\label{prt}\\
\tilde{p}_{\perp}&\equiv&p_{\perp}+p_{\perp}^{\theta},\label{ppt}
\end{eqnarray} 
as the effective physical quantities.

Now, as was shown in \cite{contrerasextended},
a decoupling in the geometric sector can be  successfully implemented through
\begin{eqnarray}\label{decoupling}
\nu&=&\xi + \alpha g\label{g1}\\
e^{-\lambda}&=&e^{-\mu}+\alpha f\label{f},
\end{eqnarray}
where $g$ and $f$ are the geometric deformation undergone by $\xi$ and $\mu$, ``controlled'' by the free parameter $\alpha$. It is important to note that Eqs. (\ref{g1}) and (\ref{f}) are not merely coordinate transformations. Indeed, they can be viewed as a consequence of a source with energy momentum tensor given by $\theta_{\mu\nu}$.
%, the system will be transformed in such a way that the equation of motions associated with the mentioned source will satisfy an effective “quasi- Einstein system”.
Now, replacing (\ref{g1}) and (\ref{f}) in  (\ref{eins1})-(\ref{eins2}) 
we obtain on the one hand 
\begin{eqnarray}
\kappa^{2}\rho &=&-\Lambda +\frac{e^{-\mu} \mu'}{2 r}\label{iso1}\\
\kappa ^2 p_{r}&=&\Lambda +\frac{e^{-\mu} \xi'}{2 \kappa ^2 r}\label{iso2}\\
\kappa ^2 p_{\perp}&=&\Lambda -\frac{e^{-\mu} \left(\mu' \xi'-2 \xi''-\xi'^2\right)}{4}
,\label{iso3}
\end{eqnarray}
with $T_{\mu\nu}^{(s)}$ satisfying,
\begin{eqnarray}
\nabla^{(\xi,\mu)}_{\nu}T^{\nu(s)}_{1}=p'+\frac{1}{2}(p+\rho) \xi'=0,
\end{eqnarray}
where $(\xi,\mu)$ means that the covariant derivative is compatible with the metric (\ref{g1}) and (\ref{f}) with $\alpha=0$. Since the metric $(\xi,\mu)$ is a solution of Einstein equations (\ref{iso1}), (\ref{iso2}) and (\ref{iso3}) which is independent of $\theta_{\mu\nu}$ we will say that they represent the
\textit{seed sector}. On the other hand, we have the set of equations of motion for the source $\theta_{\mu\nu}$, they read
\begin{eqnarray}
\kappa ^2\rho^{\theta}&=&-\frac{\alpha  f'}{2 r}\label{aniso1}\\
\kappa^{2} p_{r}^{\theta}&=&\alpha Z_{1}+\frac{\alpha f\nu'}{2 r}\label{aniso2}\\
\kappa^{2} p_{\perp}^{\theta}&=&\alpha Z_{2}+\frac{\alpha}{4}f'\nu+\frac{\alpha}{4 } 
f\left(2 \nu ''+\nu '^2\right),\label{aniso3}
\end{eqnarray}
where
\begin{eqnarray}
Z_{1}&=&\frac{e^{-\mu} g'}{2 r}\\
Z_{2}&=&\frac{1}{4} e^{-\mu} \left(2 g''+g' \left(\alpha  g'-\mu'+2 \xi'\right)\right).
\end{eqnarray}
It is necessary to mention at this point that, although it may seem surprising, the transformation performed in equations (\ref{eins1}), (\ref{eins2}) and (\ref{eins3}) which brings us to systems of equations (\ref{iso1}), (\ref{iso2}), (\ref{iso3}) and (\ref{aniso1}), (\ref{aniso2}), (\ref{aniso3}) is extensible to $3 + 1$ dimensions, which is why it represents a remarkable fact. Note that Eqs. (\ref{aniso1}), (\ref{aniso2})
and (\ref{aniso3}) correspond to a set of quasi--Einstein equations sourced by $\theta_{\mu\nu}$ satisfying
\begin{eqnarray}\label{constheta}
\nabla_{\rho}\theta^{\rho}_{\nu}=
-\frac{1}{2}\alpha g'(p+\rho)\delta_{\nu}^{1},
\end{eqnarray}
which ensures the conservation of the total energy momentum tensor. It is worth mentioning that when $\alpha=0$ the solution corresponds to that parametrized by the metric of the seed sector. Now, given that for $\alpha\ne 0$ the enegy--momentum tensor $\theta_{\mu\nu}$ induce a deformation in the metric, it is said that (\ref{aniso1}), (\ref{aniso2}) and (\ref{aniso3}) represent the equations of the
\textit{decoupling sector}. Furthermore, it is noticeable that (\ref{constheta}) means that a decoupling without energy--momentum exchange 
can be reached either by imposing $g'=0$ or $p+\rho=0$. The former requirement corresponds to the standard
MGD where only $g^{-1}_{rr}$ undergoes a geometrical deformation. The latter
entails a barotropic equation of state in the isotropic sector. What is more, if the isotropic 
sector is vacuum (the exterior of a star), the barotropic condition is trivially fulfilled and
the decoupling without exchange of energy--momentum is straightforward. We conclude this section
pointing out that the conditions for the decoupling of the sources $T^{(s)}_{\mu\nu}$ and
$\theta_{\mu\nu}$ coincides with those found for the $3+1$ dimensions case reported in Ref. {\cite{Ovalle:ext}} where is effectively reported
that both sources can be successfully decoupled as long as there is exchange of energy between them.

\section{$2+1$ Einstein--Klein--Gordon system: the minimal case}\label{EKG}

In the previous section we obtained that the decoupling sector is a system of three equations for five unknowns variables, $\{f,g,\rho^{\theta},p_{r}^{\theta},p_{\perp}^{\theta}\}$. Of course, this number decreases to four unknowns when the minimal geometric deformation is implemented, namely, when $g\to0$ which besides, from
(\ref{constheta}), implies
\begin{eqnarray}\label{theconsmgd}
\nabla_{\rho}\theta^{\rho}_{\nu}=0.
\end{eqnarray}
The meaning of (\ref{theconsmgd}) is clear: both the seed and the decoupling sectors are conserved independently so that the interaction is only gravitational. However, in order to integrate the system the imposition of an extra constraint is mandatory. Note that, if $\theta_{\mu\nu}$ remains as a generic source, any suitable equation of state can be used to reduce the number degrees of freedom. Indeed,
it was the case reported in \cite{Contreras2019} where the GD by MGD was implemented to deform a BTZ background by implementing isotropic condition ($p^{\theta}_{r}=p_{\perp}^{\theta}$), conformally flat condition ($p^{\theta}_{\perp}=\rho^{\theta}-p_{r}^{\theta}$) and a linear anisotropic equation of state ($\rho^{\theta}=a p_{r}^{\theta}+b p_{\perp}^{\theta}$).

In this work we shall follow an alternative strategy. To be more precise, we associate 
$\theta_{\mu\nu}$ with a physical field to explore if this extra constraint required to integrate the system arises naturally. To this end we assume the Einstein--Klein--Gordon system
\begin{eqnarray}\label{ekg}
S_{EKG}=\int d^{4}x \sqrt{-g}\left(\frac{R-2\Lambda}{2\kappa}-\frac{1}{2}\partial_{\mu}\Phi\partial^{\mu}\Phi-V(\Phi)\right),
\end{eqnarray}
where $\Phi$ es the scalar field and $V(\Phi)$ the self interaction potential (see \cite{Ovalle:2018ans} for implementation in $3+1$ dimensional spacetimes). From (\ref{ekg}), we can identify
\begin{eqnarray}\label{theta}
\alpha \theta_{\mu\nu}=-\frac{2}{\sqrt{-g}}\frac{\delta \mathcal{L}_{\phi}}{\delta g^{\mu\nu}} ,
\end{eqnarray}
with
\begin{eqnarray}\label{Lfi}
\mathcal{L}_{\phi}=-\sqrt{-g}\left(\frac{1}{2}\partial^{\mu}\Phi\partial_{\mu}\Phi+V(\Phi)\right).
\end{eqnarray}
Now, when the minimal deformation is impossed, (\ref{aniso1}), (\ref{aniso2}) and (\ref{aniso3}) read (using $\kappa^{2}=8\pi$)
\begin{widetext}
\begin{eqnarray}
-\frac{1}{2} e^{-\lambda } \Phi '^2-V(\Phi )&=&\frac{\alpha  f'}{16 \pi  r}\label{kg1}\\
\frac{1}{2} e^{-\lambda } \Phi '^2-V(\Phi )&=&\frac{\alpha  f \nu '}{16 \pi  r}\label{kg2}\\
-\frac{1}{2} e^{-\lambda } \Phi '^2-V(\Phi )&=&\frac{\alpha  \left(f' \nu '+f \left(2 \nu ''+2 \nu '^2\right)\right)}{32 \pi }\label{kg3}.
\end{eqnarray}
\end{widetext}
Note that $T^{\mu\nu}_{\;\;\;;\mu}=0$  leads to
\begin{eqnarray}
\Phi''+\left(\frac{2}{r}+\frac{1}{2}(\nu'-\lambda')\right)\Phi'=\frac{dV}{d\Phi}e^{\lambda},
\end{eqnarray}
which corresponds to the explicit form of the Klein--Gordon equation in circular symmetry.

Next, a direct comparison between Eqs. (\ref{kg1}) and (\ref{kg3}) reveals that
\begin{eqnarray}\label{constaint}
\frac{\alpha  f' \left(r \nu '-2\right)}{r}+2 \alpha  f \left(\nu ''+\nu '^2\right)=0,
\end{eqnarray} 
which corresponds to an extra constraint useful to close the system of equations. Note that (\ref{constaint}) constitutes
a differential equation for the decoupling function $f(r)$ which is independent of both the scalar field $\Phi$ and the self--interacting potential $V(\Phi)$. In this sense, we have 
not only provided a constraint to solve the system but we have
reduced the problem to solve (\ref{constaint})
for the decoupling function $f$ for some seed sector. However, note that given the non--linearity of (\ref{constaint}), finding analytical solutions for $f$ could be a non--trivial task. Indeed, it will depend on the functional form of the metric potential $\nu$. In the next section 
we shall provide the metric functions $\{\nu,\mu\}$ associated to BTZ and a $2+1$ dimensional BH with a Coulomb field to solve (\ref{constaint}) and finally obtain $\Phi$ and $V(\Phi)$.

\subsection{Case 1. BTZ}
In this section we assume the well--known static BTZ as the seed solution, which metric functions are given by
\begin{eqnarray}\label{BTZ}
e^{\nu}=e^{-\mu}=-M+\frac{r^{2}}{L^{2}},
\end{eqnarray}
where $L^{2}=-1/\Lambda$.
In three dimensions the full curvature tensor is defined by the Ricci tensor, so any smooth solution of Einstein’s equations is a space-time of constant curvature. This solution is asymptotically anti--de Sitter and has a horizon located at 
\begin{eqnarray}
r_{H}=L\sqrt{M},
\end{eqnarray}
which is both a Killing ($g_{tt}=0$) and a causal ($g^{rr}=0$). Replacing (\ref{BTZ}) in (\ref{constaint}) we obtain
\begin{eqnarray}\label{f1}
f=c_1 e^{-\frac{r^2}{L^2 M}}.
\end{eqnarray}
so that the total solution reads
\begin{eqnarray}
e^{\nu}&=&-M+\frac{r^{2}}{L^{2}}\label{totsol1}\\
e^{-\lambda}&=&-M+\frac{r^{2}}{L^{2}}+\alpha c_1 e^{-\frac{r^2}{L^2 M}}\label{totsol2}.
\end{eqnarray}
At this point some comments are in order. First, note that
the total solution is asymptotically anti--de Sitter. Second, its Killing horizon coincides with the BTZ solution but the causal horizon determined by the condition $e^{{\lambda(r_{c})}}=0$ is given by 
\begin{eqnarray}
\frac{r_{c}^{2}}{L^{2}}+\alpha c_1 e^{-\frac{r_{c}^2}{L^2 M}}=M,
\end{eqnarray}
which only coincides with the Killing horizon when $\alpha=0$. However, $r_{c}>r_{H}$ in general, so the signature of the metric becomes $(--+)$ for some $r_{H}<r<r_{c}$ which should be discarded. In this sense,
although the solution can not be interpreted as a proper BH, we can consider it as the exterior of a $2+1$ star with radius $R>r_{c}$.

Next, using (\ref{f1}) in (\ref{kg1}), (\ref{kg2}) and (\ref{kg3}) we obtain $\Phi$ as a formal solution
\begin{eqnarray}\label{fiformal}
\Phi=-\sqrt{\frac{\alpha c_{1}}{8\pi}}\int_{}^{} \frac{ u}{ \sqrt{L^2 M \left(u^2-L^2 M\right) \left(\alpha  c_1+e^{\frac{u^2}{L^2 M}} \mu (u)\right)}}du,
\end{eqnarray}
from where
\begin{eqnarray}\label{potencial}
V(\Phi)=\frac{\alpha  c_1  \left(2 L^2 M-r^2\right)}{16 \pi  L^2 M \left(L^2 M-r^2\right)}e^{-\frac{r^2}{L^2 M}}.
\end{eqnarray}
Note that, since an analytical solution for the scalar field can not be obtained, the interpretation of the potential as a function of $\Phi$ is not possible. However, we can go a step further as long as some appropriate limiting values of the parameters involved are considered. For example, if the integrand in 
(\ref{fiformal}) is expanded around $\alpha<<1$ and $\frac{r^{2}}{L^{2}M}<<1$, we obtain 
\begin{eqnarray}\label{campobtz}
\Phi ^2\approx\frac{\alpha  c_1}{32 \pi  M}\left(\frac{r}{L^{2}M}\right)^{4}.
\end{eqnarray}
We see that, in this limit, the scalar field depends on the value of $r^2$. This divergent behavior is not surprising, it is a known fact  that just inherits from BTZ's solution, which is asymptotically anti--de Sitter due to the remnant field located in spatial infinity. From (\ref{campobtz}) we get,
\begin{eqnarray}\label{vfi1}
V(\Phi)\approx\frac{1}{L^{4}M^{2}}\left(\frac{\alpha  c_1}{8 \pi }-\sqrt{\frac{\alpha c_1 M  }{ 8 \pi  }}\Phi+2 M \Phi ^2\right).
\end{eqnarray}
At this point some comments are in order. First, note that exponential potentials such as that of Eq. (\ref{potencial}) are typical for moduli or dilaton fields \cite{Barreiros2000}. Even more, 
the potential (\ref{vfi1}) looks formally (with the exception of a constant term) as the potential in the spatially flat (2+1) cosmological models with \cite{Garcia-Diaz:2017cpv}
\begin{eqnarray}\label{selfint}
V(\Phi) = A (\alpha \Phi^{2/(1-\beta)} -\Phi^{2\beta/(1-\beta)}).
\end{eqnarray}
Second, it is worth noticing that the first term is a constant which depend on the decoupling parameter $\alpha$. Now, it is clear that at the level of the action (\ref{ekg})
this constant modifies the values of the cosmological term $\Lambda$. More precisely, the first term of the potential leads to a effective cosmological constant which reads
\begin{eqnarray}
\Lambda_{eff}=\Lambda+\alpha\frac{\kappa c_{1}}{8\pi L^{4}M^{2}}.
\end{eqnarray}
Furthermore, as we are considering a negative cosmological constant, namely $\Lambda=-1/L^{2}$, the effect of the extra term arising from the potential will depend on the sign of $\alpha c_{1}$. In summary, the identification of the decoupling sector with the scalar field has a non--trivial effect on the cosmological constant through the decoupling parameter. Then, the
third term in (\ref{vfi1}) can be interpreted as the massive term of the scalar field as usual. Finally, given that the second term in (\ref{vfi1}) is linear in $\Phi$, it can not be interpreted as a self interacting term. It is known that polynomial expansions of scalar field potentials that include odd terms have not yet a clear physical interpretation but in this case is a direct consequence of the decoupling sector mechanism and cannot be removed. 

Despite these issues, the aforementioned linear term could be of importance in relation with dark energy. Specifically, one of the first models of dark energy \cite{Linde1986} considered to replace the cosmological constant by the energy density of a slowly changing scalar field
with a linear effective potential $V(\Phi)=V_{0}\left(1 + \beta \Phi \right)$, where $\beta$ is a parameter such that $|\beta|$ is related with the lifetime of the Universe. Even more, 
it has been argued that such a potential is favored by anthropic principle considerations \cite{Garriga2000,Garriga2003,Garriga2004,Garriga2004b} and can solve the coincidence problem \cite{Avelino2005}. Recently, the fate of a universe driven by a linear potential has been the subject of Refs. \cite{Kaloper2015,Ferreira2018} in light of sequestration of vacuum energy and the end of the universe. Finally, we note that string cosmology predicts that, in the presence of suitable wrapped branes, the potential energy grows linearly with the canonically normalized inflaton field \cite{McAllister2010}. In this sense, our model indicates that MGD could provide a natural  way of creating dark energy- or certain string-like mechanism in 2+1 dimensions.

\subsection{Case 2. $2+1$ dimensional BH with a Coulomb field}

In this case, we consider the $2+1$--dimensional BH with Coulomb field given by
\begin{eqnarray}
e^{\nu}=e^{-\mu}=-M+\frac{r^{2}}{L^{2}}+\frac{4q^{2}}{3r},
\end{eqnarray}
as the seed solution. Just like in the BTZ case, the Killing and the causal horizons coincides and are located at
\begin{eqnarray}
\frac{r_{H}^{2}}{L^{2}}+\frac{4q^{2}}{3r_{H}}=M.
\end{eqnarray}

Now, from (\ref{constaint}) the decoupling function 
is given by
\begin{eqnarray}
f=c_{1}r^{4/3}e^{-\frac{r}{L^{2}M^{2}}
	\left(4q^{2}+M\right)}(2q^{2}-M r)^{-\left(\frac{4}{3}+\frac{8q^{4}}{L^{2}M^{3}}\right)},
\end{eqnarray}
from where the total solution reads 
\begin{widetext}
\begin{eqnarray}
e^{\nu}&=&-M+\frac{r^{2}}{L^{2}}+\frac{4q^{2}}{3r}\\
e^{-\lambda}&=&
-M+\frac{r^{2}}{L^{2}}+\frac{4q^{2}}{3r}
+\alpha c_{1}r^{4/3}e^{-\frac{r}{L^{2}M^{2}}
	\left(4q^{2}+M\right)}(2q^{2}-M r)^{-\left(\frac{4}{3}+\frac{8q^{4}}{L^{2}M^{3}}\right)}
\end{eqnarray}
\end{widetext}
In contrast to the previous case, we can enforce $e^{\nu(r_{H})}=e^{-\lambda(r_{H})}$, namely
\begin{eqnarray}
\alpha=\frac{e^{\frac{r_{H} \left(M r_{H}+4 q^2\right)}{L^2 M^2}} \left(2 q^2-M r_{H}\right)^{\frac{8 q^4}{L^2 M^3}+\frac{4}{3}} \left(L^2 \left(3 M r_{H}-4 q^2\right)-3 r_{H}^3\right)}{3 c_1 L^2 r_{H}^{7/3}},
\end{eqnarray}
so that the total solution has a well defined event horizon. Note also that, $e^{-\lambda}$ has a critical point at
\begin{eqnarray}
r^{*}=\frac{2q^{2}}{M},
\end{eqnarray}
which can be hidden in the interior horizon when $r^{*}<r_{H}$.

For the expansion argued above $\alpha<<1$, $q^{2}/r<<1$ and $r^{2}/L^{2}M<<1$, the the field takes the form
\begin{eqnarray}
\Phi\approx \sqrt{\frac{\alpha  c_1}{32 \pi  M^{7/3}}}
\left(\frac{r}{L^{2}M}\right)^2 ,
\end{eqnarray}
again presenting an $r^2$-type divergence inherited from the asymptotic behavior present in the seed solution. We emphasize that this is a known fact and does not depend on our development. The potential associated to this solution reads,
\begin{widetext}
\begin{eqnarray}
V(\Phi)=\frac{1}{L^4 M^2}\left(\frac{\alpha  c_1}{8 \pi  M^{4/3}}-  \sqrt{\frac{\alpha  c_1}{8 \pi  M^{1/3}}}\Phi\right),
\end{eqnarray}
\end{widetext}
which again looks like Eq. (\ref{selfint}), including the extra constant term. As we already said, this produces a modification in the cosmological constant. Also, in this case, the potential depends on the dark-energy or string-like linear term. 

\section{$2+1$ Einstein--Klein--Gordon system: extended case}\label{ec}

In the previous section we reduced the problem of solving the Einstein--Klein--Gordon
system to seek for solutions of the constraint involving the decoupling function $f$ through the MGD approach. It is worth mentioning that solving the same problem without MGD is clearly more involved in the sense that two extra constraints are required to close the system. However,  in the framework of MGD we reduce the degrees of freedoms by providing a seed solution so the other condition arise naturally from the structure of the matter sector associated to the scalar field. Nevertheless, the original of horizon is lost given that only the $g^{rr}$ function is modified. In this regard, if we want to conserve the same horizon structure, we can use the extended version of MGD where $g\ne 0$.  In this case, Eq. (\ref{theta}) reads 
\begin{eqnarray}
-\frac{1}{2} e^{-\mu } \Phi '^2 \left(\alpha  f e^{\mu }+1\right)-V(\Phi )&=&\frac{\alpha ^2 f'}{16 \pi  r},\\
\frac{1}{2} e^{-\mu } \Phi '^2 \left(\alpha  f e^{\mu }+1\right)-V(\Phi )&=&\alpha  \left(\frac{\alpha  f \left(\alpha  g'+\xi '\right)}{16 \pi  r}+\frac{\alpha  e^{-\mu } g'}{16 \pi  r}\right),\\
-\frac{1}{2} e^{-\mu } \Phi '^2 \left(\alpha  f e^{\mu }+1\right)-V(\Phi )&=&\frac{\alpha }{32 \pi } \bigg(\alpha  \left(f' \left(\alpha  g'+\xi '\right)+f \left(2 \alpha  g''+\left(\alpha  g'+\xi '\right)^2+2 \xi ''\right)\right)\nonumber\\
&&+e^{-\mu } \left(2 \alpha  g''+\alpha  g'\left(\alpha  g'-\mu '+2 \xi '\right)\right)\bigg),
\end{eqnarray}
from where
\begin{eqnarray}
\frac{f' \left(\alpha  r g'+r \xi '-2\right)}{r}+f \left(2 \alpha  g''+\left(\alpha  g'+\xi '\right)^2+2 \xi ''\right)+e^{-\mu } \left(2 g''+g' \left(\alpha  g'-\mu '+2 \xi '\right)\right).
\end{eqnarray}
As a particular case, let us consider again the BTZ geometry as the seed solution, 
\begin{eqnarray}\label{mubtxextended}
e^{\xi}=e^{-\mu}=M-\frac{r^{2}}{L^{2}}.
\end{eqnarray}
Now, a straightforward computation reveals that, after choosing
\begin{eqnarray}\label{g}
g=\frac{2}{\alpha } \log \left(\frac{\alpha  r-2 c_1}{c_{2}}\right),
\end{eqnarray}
with $c_{1}$ and $c_{2}$ constants with dimensions of length, we obtain
\begin{eqnarray}
f=\left(-M+\frac{r^{2}}{L^{2}}\right)
\frac{4c_{1}L^{2}r^{3}-\alpha r^{6}+L^{2}c_{3}}{(\alpha r^{3}-2c_{1}L^{2}M)^{2}}=0,
\end{eqnarray}
with $c_{3}$ a constant with dimensions of a length to the power of three. It is worth mentioning that
equation (\ref{g}) is the simplest choice which allows analytical solution for $f$. Another possibility could be assuming some equations of state of the $\theta$ sector but is easy to check that the classical relations; namely, barotropic and polytropic equations, 
lead to complicated relations which cannot be integrated analytically. Now, the total solution reads
\begin{eqnarray}
e^{\nu}&=&\left(-M+\frac{r^{2}}{L^{2}}\right) \left(\frac{\alpha  r-2 c_1}{c_{2}}\right)^{2}\label{nuex},\\
e^{-\lambda}&=&\left(-M+\frac{r^{2}}{L^{2}}\right) 
\left(
1+\alpha\frac{4c_{1}L^{2}Mr^{3}+c_{3}L^{2}-\alpha r^{6}}{(\alpha  r^{3}-2c_{1}L^{2}M)^{2}}
\right).\label{lambext}
\end{eqnarray}
Note that, Eqs. (\ref{nuex}) and (\ref{lambext}) looks formally as solutions with the same horizon structure that BTZ. 

Then, with the aim to recover the BTZ solution when $\alpha\to0$, Eq. (\ref{nuex}) reveals that $c_{1}/c_{2}=-1/2$. Moreover, this condition implies that $e^{\nu}$ has no extra roots
so that the Killing horizon coincides with its BTZ counterpart. Second, to ensure that the causal horizon is the same as in the BTZ solution, we require that the factor
\begin{eqnarray}
1+\alpha\frac{4c_{1}L^{2}Mr^{3}+c_{3}L^{2}-\alpha r^{6}}{(\alpha L r^{3}-2c_{1}L^{3}M)^{2}},
\end{eqnarray}
has no real roots. This condition leads to
\begin{eqnarray}
\alpha  c_3 L^2+L^6 M^2-2 \alpha  \left(L^2-1\right) L^2 M r^3+\alpha ^2 \left(L^2-1\right) r^6\ne0,
\end{eqnarray}
which can be achieved if we choose $L=1$. Finally, from (\ref{lambext}) we must ensure that $\alpha>0$ to avoid the apparition of critical points.
For $\alpha<<1$ and $r^{2}/L^{2}M<<1$ the field takes the form
\begin{eqnarray}
\Phi^{2} \approx\frac{\alpha ^2}{4 \pi  }r, 
\end{eqnarray}
and the potential
\begin{widetext}
\begin{eqnarray}
V(\Phi)\approx\frac{\alpha ^5 M}{128 \pi ^2  \Phi ^2 \left(2 \pi  \Phi ^2-\alpha \right)}.
\end{eqnarray}
\end{widetext}
We observe that this potential is regular whenever $\Phi\ne \sqrt{\alpha/2\pi}$. Besides, as it can not be written in terms of powers of $\Phi$ its interpretation
is less straightforward in comparison with the previous cases under consideration. In any case, we have obtained, for the same limit considered previously, an analytical expression for the self-interaction potential using the extended MGD method, which in itself provides an advantage in the sense that it represents a more general method. The case of the Coulomb field was considered but without the feasibility of finding analytical solutions it does not represent such an interesting result.

Before concluding this section we would like to 
discuss some 
aspects about the asymptotic behaviour of the solutions (\ref{nuex}) and (\ref{lambext}). Note that, although $g_{rr}$ 
is asymptotically BTZ as $r\to\infty$, the $tt$ component has two extra terms which violates the desired limit. This issue could be addressed if we consider that the geometry described by (\ref{nuex}) and (\ref{lambext}), which we shall call $\mathcal{M}_{1}$, must be matched
with another geometry $\mathcal{M}_{2}$ (the BTZ solution for example) though the Darmois conditions. In this sense, the total manifold $\mathcal{M}$ corresponds the union $\mathcal{M}_{1}\cup \mathcal{M}_{2}$. The above strategy has been broadly used to construct, for example, vacuum bubbles characterized by having an inner vacuum region separated by a thin layer of matter from an outer region in the context of black holes, wormholes solutions
(see for example \cite{ei1,ei2}, and references there in).

The other possibility is to set the free parameters in order to avoid such a undesired behaviour as we shall work out in what follows. First, let us consider $c_{2}=a(1-b)^{-1}$ and $c_{1}=-a(1+\alpha)^{1/2}(1-b)^{-1}/2$ with $a$ a constant with dimensions of length and $b$ is a dimensionless constant. With this choice, Eq. (\ref{nuex}) reads,
\begin{eqnarray}
e^{\nu}=\left(-M+\frac{r^{2}}{L^{2}}\right)\left(\frac{\alpha(1-b)r}{a}+(1+\alpha)^{1/2}\right)^{2},
\end{eqnarray}
which ensures the desired asymptotic behaviour when $b\to1$, namely
\begin{eqnarray}
e^{\nu}=\left(-M+\frac{r^{2}}{L^{2}}\right)\left(1+\alpha\right)
\end{eqnarray}

Besides, the original BTZ solution is recovered when $\alpha\to0$ as required. In the case of Eq. (\ref{lambext}) we obtain two solutions depending on the choice of $c_{3}$.  
For example, if we replace the expressions of $c_{1}$ and $c_{2}$ previously defined, but leaving $c_{3}$ arbitrary we obtain that (\ref{lambext}) reduces to $e^{-\mu}$ in
(\ref{mubtxextended}). However, by taken $c_{3}=a^4 (\alpha +1)(1-b)^{-2}$, Eq. (\ref{lambext}) reads
\begin{eqnarray}
e^{-\lambda}=\left(-M+\frac{r^{2}}{L^{2}}\right)\left(
1+\frac{\alpha a^{2}}{L^{2}M^{2}}\right),
\end{eqnarray}
which reduces to the BTZ metric when $\alpha\to 0$ as required. However, in this case the scalar field and its corresponding self--interaction potential are trivial. Indeed, an explicit computation reveals that
\begin{eqnarray}
\Phi&=&\Phi_{0}=\textnormal{constant}\\
V(\Phi)&=&-\frac{a^2 \alpha ^2}{8 \pi  L^4 M^2}=\textnormal{constant}.
\end{eqnarray}

\section{Conclusions}

The existence of stable and regular solutions to the Einstein--Klein--Gordon equations suggest that self-gravitating objects formed by scalar fields can be astrophysically relevant. The study of the gravitational interaction between several objects of this type with other sources could be essential in many phenomena, so, in this sense, any related study may be useful. In this work we have implemented the Gravitational Decoupling in the framework of the Minimal Geometric Deformation approach to the Einstein--Klein--Gordon system in $2+1$ dimensional spacetimes. We found that after identifying the decoupling sector with the energy--momentum associated to the scalar field, the system of differential equations lead to a constraint which allows to obtain the decoupling function. Besides, the method allows to obtain analytical solutions for the scalar field and the self--interacting potential after considering some limiting values of the parameters involved performing the expansion around $\alpha<<1$ and $r^{2}/L^{2}M<<1$. In this sense, we conclude that the gravitational decoupling can be used to solve, at least to some extent, the Einsten-Klein--Gordon system without assuming any ansatz for the self interacting potential as it is usually done in the literature.\\

In order to have a more complete view of the problem, a different seed solution could be considered in order to achieve the gravitational decoupling. Then, it would be interesting to compare the results obtained for the scalar fields and their respective self-interaction potentials in order to obtain more clear interpretations. A possible generalization is to use the complete BTZ solution \cite{Banados1992} (taking into account symmetry considerations over the metric) containing now two integration constants, $M$ and $J$, the mass and angular momentum of the black hole. This solution could be of interest since it also has two Killing vectors (instead of one) associated with their respective, space-time, symmetries.\\

Another possible extension of the present work may consist in considering more general couplings and possibly including other fields in the treatment. It can be studied Einstein–dilaton solutions in (2 + 1) gravity or even extend it to the Einstein–Maxwell–dilaton case. There, the Schwarzschild coordinate frame is used to determine static cyclic symmetric metrics for (2 + 1) Einstein equations coupled to an electric Maxwell field and a dilaton in the presence of an exponential potential. The general solution can be derived and identified with the Chan–Mann charged dilaton solution \cite{CHM2}. In this sense a family of stationary dilaton solutions can been generated; the solutions possessing a richer variety of parameters: dilaton and cosmological constants, charge, momentum, and mass. These solutions have been characterized by their quasi-local energy, mass, and momentum by means of expansions at spatial infinity. This fact could be interesting to connect with the interpretation of the scalar potential. These and other considerations can and will be addressed in future research.

\section*{ACKNOWLEDGEMENTS}
P. B. is funded by the Beatriz Galindo contract BEAGAL 18/00207 (Spain).

%%%%%%%%%%%%%%%%%%%%%%%%%%%%%%%%%%%%%%%%%%%%%%%%%%%%%%%%%%%%%%%%%%%%%%%%%%%%%%%%%%%%%%%%%%%%%%%%

\end{document}